\pdfoutput=1

\documentclass[sigconf,authorversion,nonacm]{acmart}

\usepackage{color}

\AtBeginDocument{%
  \providecommand\BibTeX{{%
    \normalfont B\kern-0.5em{\scshape i\kern-0.25em b}\kern-0.8em\TeX}}}

\begin{document}

\title{The Many Facets of Trust in AI: Formalizing the Relation Between Trust and Fairness, Accountability, and Transparency}
\author{Bran Knowles}
\email{b.h.knowles1@lancaster.ac.uk}
\affiliation{%
  \institution{Lancaster University}
  \city{Lancaster}
  \country{UK}
}

\author{John T.\,Richards}
\email{ajtr@us.ibm.com}
\affiliation{%
  \institution{TJ Watson Research Center, IBM}
  \city{Yorktown Heights}
  \state{New York}
  \country{USA}
}

\author{Frens Kroeger}
\email{frens.kroeger@coventry.ac.uk}
\affiliation{%
  \institution{Coventry University}
  \city{Coventry}
  \country{UK}
}

\renewcommand{\shortauthors}{Knowles, Richards \& Kroeger \copyright August 2022}

\begin{abstract}
Efforts to promote fairness, accountability, and transparency are assumed to be critical in fostering Trust in AI (TAI), but extant literature is frustratingly vague regarding this ``trust''. The lack of exposition on trust itself suggests that trust is commonly understood, uncomplicated, or even uninteresting. But is it? Our analysis of TAI publications reveals numerous orientations which differ in terms of who is doing the trusting (\textit{agent}), in what (\textit{object}), on the basis of what (\textit{basis}), in order to what (\textit{objective}), and why (\textit{impact}). We develop an ontology that encapsulates these key axes of difference to a) illuminate seeming inconsistencies across the literature and b) more effectively manage a dizzying number of TAI considerations. We then reflect this ontology through a corpus of publications exploring fairness, accountability, and transparency to examine the variety of ways that TAI is considered within and between these approaches to promoting trust.
\end{abstract}





\maketitle

\section{Introduction}

The matter of trust is strongly associated with considerations of fairness, accountability, and transparency (FAT), with trust often construed as an overarching aim that these contribute toward. This relationship is perhaps most clearly represented by the guidelines produced by The European Commission's High-Level Expert Group on AI, which identifies fairness, accountability, and transparency as three of its seven key requirements for ``Trustworthy AI'' \cite{ai2019high}. Similarly, the Montréal Declaration for Responsible AI states the collective impact of realizing its enumerated principles (which allude to FAT though adopting slightly different terminology) as ``lay[ing] the foundation for cultivating social trust toward artificially intelligent systems'' \cite{montreal}---their accompanying report \cite{montrealreport} using the word ``trust'' over 40 times. In their review of documents containing ethical principles for AI, Jobin et al.\,\cite{jobin2019global} identify trust as one of the 11 common principles (featured in 28 out of 84 documents) alongside transparency, fairness, and responsibility/accountability; and specifically found that 12 of these documents viewed transparency as key to fostering trust.

Despite consensus that Trust in Artificial Intelligence (TAI) matters, precious little research has focused specifically on TAI; and the little that does exist typically does not do justice to trust. Research typically assumes that the technology which trust relates to is complex, but trust itself is not; that the term ``AI'' unhelpfully collapses a wide range of disparate systems, but the term ``trust'' is understood the same by all. Consequently, trust often remains under- or even altogether undefined, and its dimensions and complexities are overlooked. Fundamental questions remain regarding ``what trust is, and how it can be built and maintained'' \cite{sutrop2019should}. This paper endeavours to begin closing this gap. 

In what follows, we explicate the myriad facets of TAI explored across the literature. In section 2, we analyze an established corpus of TAI publications and uncollapse the manifold ways in which trust and distrust affect the perception and use of AI. The resulting ontology reveals the full richness and import of TAI, highlighting synergies and tensions between various orientations. In section 3, we demonstrate the usefulness of the ontology by reflecting it through a second corpus of FAT publications, using it to expose under-developed dimensions of TAI and diagnose critical disconnects in how researchers justify their TAI efforts. 

\smallskip

\textit{Contributions: }
\begin{enumerate}
    \item We develop a common frame of reference for explicating different approaches to the multifaceted challenge of TAI to promote greater clarity in the discourse.
    \item We identify core pillars representing prominent and relatively coherent TAI narratives that may be useful for orienting ones own approach to TAI.
    \item We demonstrate the further usefulness of this common frame by applying it to FAT literature to reveal important differences in conceptions of TAI underlying the three dominant approaches to trust-promotion, as well as potential anomalies deserving of attention.
\end{enumerate}

\section{Surveying the field}

The first corpus of texts underlying our review of trust in AI literature was created through a moderated process of selection and group discussion. The working group tasked with compiling a list of texts relevant to the topic of trust in AI was assembled by the Partnership on AI (PAI) 
(see the acknowledgements section at the end of this paper).\footnote{While the conclusions about the state of the literature to date presented in what follows  owe a number of impulses to the discussions of the working group, as summarized in a Key Insights document (available: https://tinyurl.com/3pwfjpux), here the authors have tried to go much further in depth of analysis regarding axes of difference within the literature.} It consisted of 30 industry professionals and academics who had significant expertise relevant to the topic of trust in AI, with a more active core group consisting of 10 individuals including the authors of this paper. Working group members were asked to submit their suggestions for relevant articles to an online repository, which was subsequently shared with the group. Members were then asked to highlight texts on whose inclusion in the repository they disagreed; texts were removed if at least two working group members had recommended this independently of each other. After the removal of these texts as well as duplicates, the repository comprised 78 texts, mostly academic in nature, but also including a number of industry reports and some items on the topic which were slightly more journalistic in style. Regular catch-ups and discussions were scheduled for the working group via Zoom. Initial discussions focused on understanding the nature and composition of the corpus of texts assembled. One important topic of discussion was apparent bias within the corpus towards individual concepts (notably: explainability) and whether or not this bias should be corrected. It was decided to expand the corpus further in order to fill in several blind spots which had been identified in group discussions (e.g.\,conceptual treatments of trust stemming from social science).\footnote{As a result of deliberately enhancing the corpus with theorizations on trust, not all papers mention AI.}

The authors of this paper were given the task of assigning codes to the papers in this first corpus, principally based on their abstracts, which were designed to capture whether a paper was motivated by concerns relating to \textit{what trust means} in the context of AI, including different dimensions of trust (a category we named ``understanding''; 78\% of the corpus); the specificities of different \textit{objects of trust} in this field (``receiving''; e.g.,\,trust in data vs.\,trust in an algorithm vs.\,trust in the company owning and deploying the algorithm; 37\%); the different \textit{antecedents of and/or interventions} affecting trust or distrust  (``promoting''; 64\%); and/or the different \textit{consequences} that trust and distrust may have (``impacting''; 16\%). While these codes were developed and assigned for a purpose separate from this paper---an annotated bibliography has been made available by the Partnership on AI \footnote{Available at: https://tinyurl.com/25vjatdf.}---the exercise revealed challenges in unifying a diverse set of social phenomena referred to as `trust' \cite{bigley1998straining,dutton2006trust,mcknight2000trust,mcknight2001trust,rousseau1998not,sabater2005review,ullman2018does} and the casual appropriation of these myriad phenomena within TAI.

In order to more formally account for the differences we were observing, we carried out a two-stage qualitative content analysis \cite{hsieh2005three}. Emulating \textit{conventional content analysis}, stage one developed categories through a process of induction, resulting in key axes of difference organized around the questions: \textit{Who is doing the trusting?} \ldots \textit{in what?} \ldots \textit{on the basis of what?} \ldots \textit{in order to what?} \ldots \textit{and why?} Categories (i.e.\,inferred responses to each of these questions) were arranged into an initial web structure, which was used to guide the directed content analysis of the next stage. Stage 2 used a \textit{directed content analysis} approach \cite{hsieh2005three}: papers' orientation to ``trust in AI'' was analyzed with respect to the \textit{agent} doing the trusting, the \textit{object} of the trust, the \textit{basis} for trusting, the \textit{objective} (e.g.\,to gain, maintain, or restore trust), and the proposed \textit{impact} of having that trust, corresponding to the organizing questions above.\footnote{Interestingly for this corpus on ``Trust in AI'', not all papers specifically mention trust. For these publications, the reason for its inclusion and how it speaks to the matter of trust was inferred. This is consistent with our chosen methodology, which is concerned with differences in meanings rather than the number of instances per category.}  The aim of this exercise was three-fold: 1) to identify exemplars of each of these categories, 2) to validate and/or extend this emerging ontology to reflect the full diversity comprised by the corpus, and 3) to look for relationships between categories, specifically whether there were identifiable pathways through the diagram which might represent shared orientations within the corpus. The findings of stage 2 are reported below; the final iteration of the ontology is depicted in Figure 1, which for completeness includes an additional category, \textit{promotion}, i.e.\,the proposed mechanisms for promoting trust (see section 2.6). 

\begin{figure*}
  \includegraphics[width=.62\textwidth]{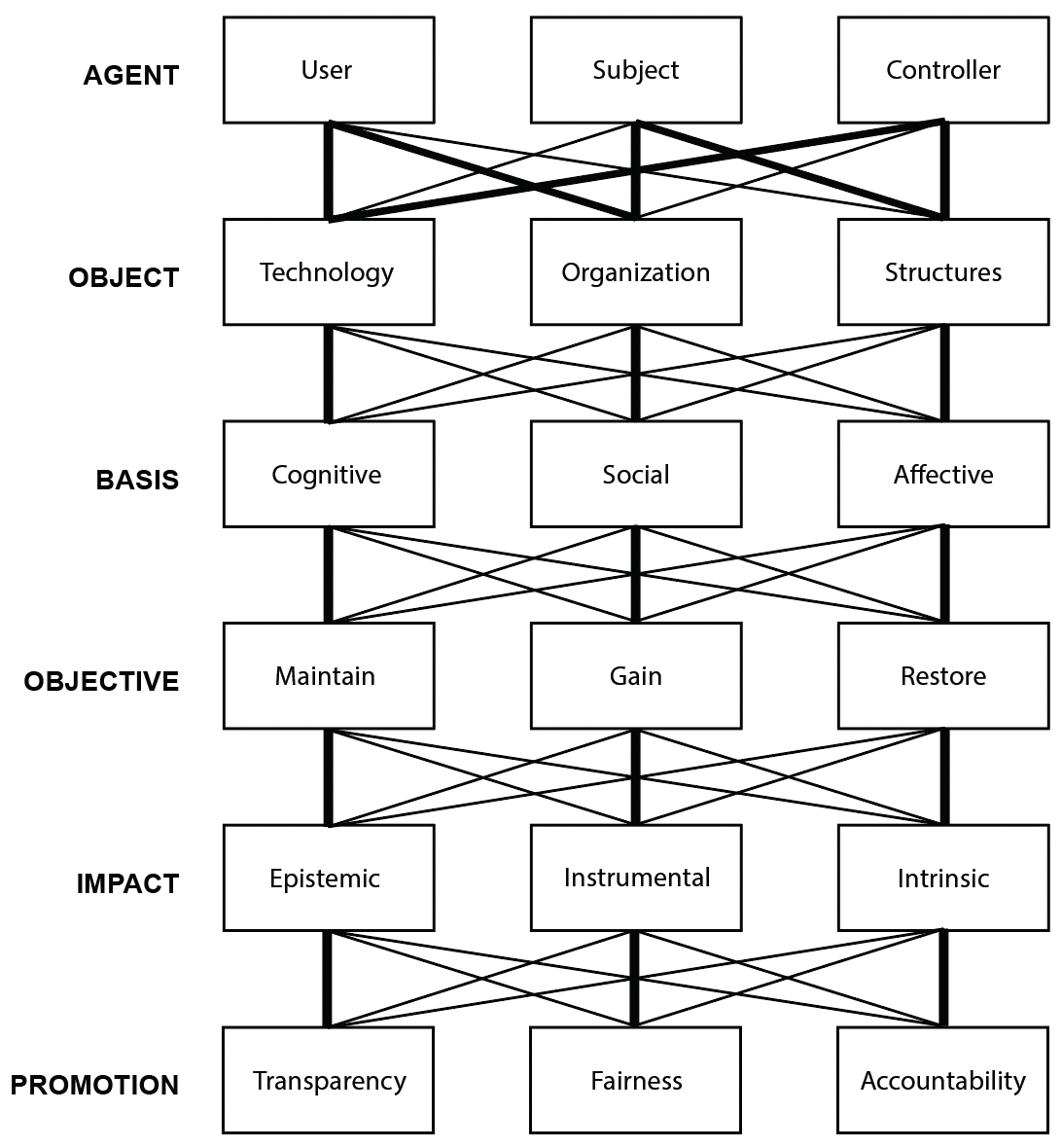}
  \caption{Ontology of orientations within the TAI literature. 
  Bold pathways indicate logical affinity between categories.}
\end{figure*}

\subsection{Agent: Who is doing the (dis)trusting?}

We identify three broad types of trustor. The first (and dominant category) are the \textit{users} who knowingly interact with AI during the accomplishment of a task, such as decision making \cite{lepri2018fair,7349687,mittelstadt2019explaining,ribeiro2016should,rieder2016datatrust}. 
These interactions raise a variety of concerns helpfully taxonomized as ``use, misuse, disuse and abuse'' of the AI \cite{parasuraman1997humans}. Explorations of use reveal ongoing ``re-negotiation'' of trust in the AI ``through practices of skepticism, assessment, and credibility'' \cite{Passi:2018:TDS:3290265.3274405}. Misuse involves the use of otherwise trustworthy AI in ways that make its effects untrustworthy, in some cases leading to catastrophic failures and complete erosion of trust (see \cite{kroeger2015development}). Misuse can result from abuse, i.e.\,implementing AI 
``without regard for the consequences for human performance'' \cite{parasuraman1997humans}, and can take the form of ``neglect'' \cite{hancock2011meta}. Often misuse occurs as a result of over-trusting \cite{Wagner:2018:ORA:3271489.3241365} or over-relying on \cite{dzindolet2003role,7349687,lee2004trust} the AI. Interestingly, publications in the corpus present evidence of \textit{automation bias}, whereby people tend to defer to machines in instances of doubt over conflicting information \cite{Wagner:2018:ORA:3271489.3241365}, as well as the contradictory impulse of \textit{algorithm aversion}, whereby ``people more quickly lose confidence in algorithmic than human forecasters after seeing them make the same mistake'' \cite{dietvorst2015algorithm}. 
Explanations for this seeming incongruity include differences in agents' predispositions (including both trusting stance \cite{hancock2011meta,mcknight2001trust,toure2015or} and risk tolerance  \cite{dutton2006trust}); membership in an affected group \cite{Woodruff:2018:QEP:3173574.3174230}; and level of AI expertise, competence, or attentional capacity \cite{hancock2011meta,siau2018building}.\footnote{We note the particularly wide spectrum of user types explored across the TAI literature---from non-expert to expert, interacting with the AI in low- to high-risk contexts. These differences matter and require careful consideration.} 
Other literature is concerned with the potential for agents to disuse the AI if believed to be insufficiently trustworthy \cite{lipton2017doctor}, with one means of preventing this being to assist users in predicting when a system might fail \cite{mittelstadt2019explaining,ribeiro2016should} (consistent with strategies for maintaining trust between stakeholders \cite{Knowles:2014:TD:2531602.2531699,7676170}). 
Some of the literature suggests that when the AI behaves in ways that do not meet users expectations, some forms of transparency can mitigate loss of trust \cite{kizilcec2016much} and empower more effective use of these systems \cite{Rader:2018:EMS:3173574.3173677}.

In contrast to the above, the \textit{subject} is generally not using AI with intentionality, nor are they in a position to disuse it. Rather, this agent is affected by the AI's recommendations \cite{Binns:2018:RHP:3173574.3173951} or wider impacts on the society in which they live. Often the subject has limited understanding of AI, but may perceive the AI as generally untrustworthy and experience high levels of anxiety about its impacts \cite{cave2019scary}.  These feelings may stem from subjects' comparative lack of agency and disproportional benefits accruing to entities who own the AI  \cite{kennedy2015data}. Subjects may seek to redress this imbalance by restricting access to their personal data for use by these systems \cite{woolley2019trust}. To understand and/or avoid unfavorable decisions being made about them by specific AIs, subjects are thought to be interested less in the `how' of the AI and more the `why'
\cite{wachter2017counterfactual}, but may face significant challenges in contesting these decisions  \cite{wachter2018gdpr}.

Finally, the \textit{controller} is concerned with assuring the trustworthiness of the AI, either as the entity developing \cite{ribeiro2016should}, purchasing (see \cite{hind2018increasing}, omitted from corpus), monitoring/assessing \cite{lepri2017tyranny}, or regulating AI (alluded to in but not a particular focus of this corpus). It is important to recognize that these types of controllers can have quite different agendas and capabilities. We have grouped these controllers together because they all determine the characteristics of the AI in a substantial sense, and do so through the application of formalized trustworthiness criteria  \cite{doshi2017towards,liu2018delayed}. Problematically, these may not map neatly to criteria \textit{subjects} or \textit{users} apply in evaluating AI's trustworthiness. Lay notions of fairness, for example, are not always aligned with statistical fairness constructs \cite{Binns:2018:RHP:3173574.3173951,binns2018fairness}; moreover, there is an important temporal dimension to fairness requiring ongoing assessment of impacts \cite{liu2018delayed}. In addition to efforts to bring these into closer alignment \cite{lee2017algorithmic,lepri2018fair} (see ``goal congruence'' \cite{siau2018building}), there is a recognition of a need to bridge the divide between regulators' and independent auditors' trustworthiness criteria and that of developers and purchasers \cite{Veale:2018:FAD:3173574.3174014}.

\subsection{Object: In what is the agent (dis)trusting?}

Most often, the object of trust within TAI is the \textit{technology} itself. Setting aside the enormous variety of AIs, there are different characteristics of the technology that can be (dis)trusted, including ``(1) the performance of the technology, (2) its process/attributes, and (3) its purpose'' \cite{siau2018building}---roughly relating to how well it works from the perspective of the user \cite{Dudley:2018:RUI:3232718.3185517,Sinha:2002:RTR:506443.506619}, how it works \cite{Rader:2018:EMS:3173574.3173677,ribeiro2016should}, and what it does \cite{balaram2018artificial,lepri2018fair}. A key route to trust in the technology is through affecting users' perceptions of performance through model interpretability \cite{chuang2012interpretation,doshi2017towards,lipton2016mythos,narayanan2018humans}. Explanations can be provided to compensate for interpretability shortcomings \cite{glass2008toward,kulesza2013too,ribeiro2016should,yin2019understanding}, though do not always lead to greater trust \cite{cramer2008effects}, e.g.\,if users are given too much information \cite{kizilcec2016much} or are not given the right kind of information \cite{7349687,Lim:2009:WWE:1518701.1519023}.  Explanations necessarily elide certain details about the model's inner workings which may prove important for being able to appropriately calibrate trust \cite{mittelstadt2019explaining}. Counterfactual explanations, which convey why one thing happened rather than another, have been proposed as more useful to enabling subjects to act, above and beyond (or without need of) understanding how the AI works \cite{miller2018explanation,wachter2017counterfactual}. Understanding how the AI works can, however, be useful for correcting mistakes \cite{kulesza2013too}, and in some cases individuals may wish to modify the algorithm itself to enhance their trust in it \cite{dietvorst2016overcoming}, or defer to humans regarding ``human aspects of context'' \cite{bellotti2001intelligibility} (also \cite{lustig2015algorithmic}) when the logical outcome is not necessarily the (morally) ``right'' outcome \cite{murphy2009beyond}. Whether or not the technology is functionally trustworthy, as in it can be trusted to work as intended, questions of trust might arise with respect to the application of that technology (``its purpose''), e.g.\,the trustworthiness of AI for automating driving \cite{waytz2014mind}, for medical decision making \cite{lipton2017doctor}, or indeed for any role formerly performed by human actors \cite{lee2018understanding}. An especially important concern is whether the AI can be trusted to promote justice \cite{Binns:2018:RHP:3173574.3173951}, and an emphasis on understanding ``the mechanics of AI'' may in fact distract from or confuse deliberations on ``the purposes to which AI can be put'' \cite{harper2019role} (postdating corpus).

A number of publications suggest a complicated intermingling of trust in the \textit{technology} and trust in the \textit{organization} \cite{ullman2018does}. Trust in AI is not simply a proxy for trust in the people who developed the AI \cite{marsh2020thinking}; however, trust in AI depends in part or in whole on the perceived trustworthiness of the organization developing or deploying the AI \cite{hengstler2016applied}, and 
the loss of trust in the algorithm has been found to lead to loss of trust in the organization behind it \cite{Woodruff:2018:QEP:3173574.3174230}. This suggests the applicability of models of trust originating in organizational studies (a number are included in this corpus \cite{bigley1998straining,kramer1999trust,lewicki2006models,rousseau1998not,schoorman2007integrative,vanneste2014trust}), though the aforementioned intermingling of objects of trust necessitates caution in simply appropriating these without careful nuancing. The TAI literature shows that organizational dynamics, such as norms and expectations and relationships with coworkers (a.k.a.\,``environmental factors'' relating to team collaboration \cite{hancock2011meta}), can influence reliance on AI outputs \cite{lee2004trust}; and that the use of AI in organizational contexts also introduces dynamics of ``meta-trust'', or ``trust a person has that the other person's trust in the automation is appropriate''~\cite{lee2004trust}. While we believe trust in the \textit{organization} is a useful category, we note that there may be quite important distinctions between types of organizations that influence the nature or quality of trust agents have in them. In particular, in contrast to private sector organizations developing or deploying AI, there is an expectation that public entities are obligated to protect individuals from harm \cite{Veale:2018:FAD:3173574.3174014}---particularly as the agent is typically a \textit{subject} of public sector AI rather than a  \textit{user}. 
 
Lastly, it can be \textit{structures} that one trusts, or the wider institution comprised of these structures (see ``institutional trust'' \cite{abdul2018trends,mcknight2001trust}).\footnote{Institutional trust is important in complex societies where the formation of interpersonal trust is impractical (see \cite{kroeger2017facework}, not in corpus), though it can produce less stable trust \cite{kroeger2015development}.} 
Structures relate to so-called ``deterrence-based trust'' \cite{rousseau1998not}, i.e.\,trust that laws and sanctions are sufficient to ensure AI's trustworthiness.\footnote{Sanctions can be applied by peers, as in the case of reputation in e-commerce \cite{sabater2005review}.} 
Public anxiety about AI is linked with a perception that those deploying AI (whether private or public entities) operate without restraint \cite{cave2018motivations}; thus a key question of structures is the strength of oversight mechanisms \cite{woolley2019trust} and whether AI can be held accountable to legislative controls placed on it \cite{bellotti2001intelligibility,banavar2016learning}. This ``macroscopic  societal  accountability'' is not satisfied through well-crafted explanations of single AIs \cite{abdul2018trends}, but rather through building trust in broader social systems. It may be an indicator of structures in need of trust-building that reliance on big data for decision making \cite{lepri2017tyranny,kennedy2015data} is more prominent in cultures where institutional trust has eroded (also see \cite{Passi:2018:TDS:3290265.3274405}); and that users (e.g.\,of Bitcoin) ``prefer algorithmic authority to the authority of conventional institutions, which they see as untrustworthy'' \cite{lustig2015algorithmic} (a finding supported by \cite{sas2017design}).

\subsection{Basis: Forming (dis)trust on the basis of what?}
TAI largely prioritizes a \textit{cognitive} view of trust that emphasizes understanding as the principal mechanism of trust-formation. Transparency, interpretability, and explanation are direct appeals to this cognitive mode; however, certain cognitive limitations may hinder these approaches. For one, these approaches can place unreasonable demands on the explainee's time and energy  \cite{bunt2012explanations,glass2008toward,7676170,kulesza2013too,narayanan2018humans}. Assuming high motivation to understand the information, however, agents may yet be limited in their ability to understand or make use of it  \cite{ananny2018seeing,Rader:2018:EMS:3173574.3173677,wachter2017counterfactual,yin2019understanding}---explanations are a social interaction, requiring that the explainer understands and responds to the beliefs and needs of the explainee \cite{miller2018explanation}. And even for the most willing and able agent, it is not possible to see what AI does at all times \cite{ananny2018seeing}.

Importantly, a ``too narrowly cognitive'' view also neglects ``emotional and social influences on trust decisions'' \cite{kramer1999trust}.
\textit{Social} (``relational'' \cite{rousseau1998not} or ``interpersonal'' \cite{lewicki2006models}) trust arises through interactions which reveal the trustworthiness of actors implicated in the development, deployment, or use of the AI. For example, people have been found to engage with Facebook as a ``quasi-person'' on the basis of conceptually linked technology and interpersonal characteristics: ``competence-functionality, integrity-reliability, and benevolence-helpfulness'' \cite{Lankton:2011:MTF:1989098.1989101}.  Much of the TAI literature draws from the legacy of the well-known Ability-Benevolence-Integrity model \cite{schoorman2007integrative} in articulating key concerns underlying social trust (e.g.\,\cite{Lankton:2011:MTF:1989098.1989101,Pu:2006:TBE:1111449.1111475,ullman2018does}; also \cite{toreini2020relationship}, not in corpus); but \textit{organizational} dynamics can also affect trust, as subjective assessments of interpersonal reliability play out within the  context in which the AI is deployed \cite{hancock2011meta,lee2004trust} (see also reflections on agency in fostering trust \cite{corbett2018going,Knowles:2015:MPT:2675133.2675154}). Attending to the latter, TAI can draw from the fact that people behave more trustworthy when they know they are going to be held accountable \cite{Erickson:2000:STA:344949.345004,Knowles:2014:TD:2531602.2531699,riegelsberger2005mechanics}. Interestingly, Interactive Machine Learning offers an opportunity for users to form a more inter-personal relationship with the AI itself, with each party affecting the behavior of the other  \cite{Dudley:2018:RUI:3232718.3185517}.

In addition to rational bases for trusting---whether \textit{cognitive} or \textit{social} reasons for believing in the object's trustworthi-\linebreak ness---there can be an \textit{affective} dimension ``grounded in reciprocated interpersonal care and concern'' which moderates trust formation (\cite{mcallister1995affect}, not in corpus). This \textit{affective} trust relates to the ``emotional bond among all those who participate in the relationship'' and the feelings associated with betrayal of trust \cite{lewis1985trust}. While this could certainly apply to \textit{organizational} objects of trust, it is an interesting question whether, or to what extent, this applies to the human-AI (\textit{technology}) trust dynamic, as there is no presumption of social exchange in the technological relationship \cite{lee2004trust}, nor the reasonable expectation that this bond will compel AI to act more trustworthy \cite{lewis1985trust}. 
People are often thrust into relationships with AI which force them to build trust in ways they wouldn't normally do socially \cite{lee2004trust}, though susceptibility to trusting anthropomorphized AI reveals a desire for individuals to form a familiar social bond \cite{Leong:2019:REW:3287560.3287591,waytz2014mind} despite the inherent social limitations of AI \cite{murphy2009beyond}. Then again, this bias is greatly dependent on whether the agent believes humans are inherently trustworthy as compared with technology 
\cite{toure2015or}, and in which situations \cite{lee2004trust,lee2018understanding}. Still, 
the literature finds evidence that agents respond positively to perceptions that the AI is ``honest'' \cite{Sinha:2002:RTR:506443.506619} and respects their autonomy \cite{glass2008toward}; and negatively when the system betrays their expectations and/or annoys them \cite{glass2008toward}; and representation can be altered to manipulate affective responses \cite{siau2018building}, as with anthropomorphization. When people have limited rational basis for trusting (e.g.\,lack of access to evidence of trustworthiness, or difficulty comprehending this evidence), the \textit{affective} component may play a greater role in making, or not making, the leap to trust (see \cite{mollering2001nature}). This is evident in the public's aversion to ``scary'' robots \cite{cave2018motivations} and fear of myriad consequences of AI \cite{balaram2018artificial}. When trusting in \textit{structures},  perceptions of situational normality (see also ``presentational base'' \cite{lewis1985trust}), or the general sense that ``everything is in the proper order'' \cite{siau2018building}, is conducive to trust. 

\subsection{Objective: What is the trust goal?} 

AI is a new addition to the cultural landscape which does not benefit (nor should it) from presumptions of trustworthiness generally afforded to more established social institutions (see \cite{lewis1985trust}). Many agents, therefore, are in the initial trust-forming stage of their relationship with AI \cite{abdul2018trends}, and their first impressions of performance and process need to be positive to \textit{gain} trust \cite{siau2018building}. Gaining trust is also an objective as it relates to agents who, for whatever reason, begin with low levels of trust  \cite{hengstler2016applied}. Typically, trust is gained slowly through direct experience of and continual reassessment of trustworthiness
\cite{lewicki2006models,Passi:2018:TDS:3290265.3274405,schoorman2007integrative,vanneste2014trust}. While this process works well for \textit{users}, who can respond to encounters with decisions and explanations provided about those decisions and re-calibrate trust accordingly, \textit{subjects} are often not even aware of their interactions with AI and may only become aware of the AI when it negatively impacts them (see \cite{knowles2021sanction}, postdating corpus); though one proposed means of gaining subject trust is through public engagement (e.g.\,``citizens juries'' to elicit the public's thoughts on ethical or unethical uses of AI) \cite{balaram2018artificial}. 

Different mechanisms are involved in \textit{maintaining} trust (see also ``continuous trust'' \cite{siau2018building}). Maintaining trust involves continuing to nurture trust while simultaneously protecting against the loss of trust (particularly catastrophic loss of trust) \cite{Knowles:2015:MPT:2675133.2675154}.  Interpretability and explainability research tends to be in service of intervening within an interaction that might otherwise reduce a user's trust (e.g.\,\cite{Lim:2009:ADI:1620545.1620576,Lim:2009:WWE:1518701.1519023,kulesza2013too}). As noted previously, maintaining trust in the \textit{technology} is particularly difficult given people's intolerance of mistakes made by AI \cite{dietvorst2015algorithm} and issues with users being able to sustain interest in explanations \cite{bunt2012explanations}. 

The objective of \textit{restoring} trust when it is lost may yet again entail different tactics, principally because the agent's stance may be characterized as distrust \cite{bigley1998straining,mcknight2001trust}. For example, restoring \textit{subject} trust in AI would involve active dismantling of culturally ingrained narratives and replacing them with more positive or neutral ones  \cite{cave2019scary,siau2018building}. Some scholars conceive of distrust as being a state of lower trust than that brought to initial encounters  \cite{schoorman2007integrative} (i.e.\,in comparison to the \textit{gain} condition), meaning that greater effort is required to raise trust to the desired level.
But distrust is more than just lack of trust; it is an active stance, making it much more difficult to recover from (see strand of literature on trust repair, e.g.\,\cite{bachmann2015repairing}, not included in corpus). Distrust often results in disuse, which tends to result in fewer subsequent trust-building experiences (see \cite{bigley1998straining}). 
In addition, distrust tends to be associated with negative \textit{affect} arising from trust having been violated \cite{kramer1999trust}, as well as the entrenchment of suspicion \cite{kramer1999trust}, and these are not easily resolved through provision of evidence of trustworthiness.

\subsection{Impact: What are the benefits of having trust?}
The logic underlying the \textit{epistemic} rationale for the importance of trust in AI is that productive synergy between \textit{user} and AI arises when trust aligns with correct beliefs regarding trustworthiness \cite{7349687,lee2004trust,Wagner:2018:ORA:3271489.3241365}. Helping the agent to form a correct (or at least interactionally beneficial \cite{adar2013benevolent}) mental model is key to ensuring an effective dynamic based on appropriate  reliance \cite{bunt2012explanations,dzindolet2003role,hancock2011meta,lipton2016mythos,Rader:2018:EMS:3173574.3173677,ribeiro2016should}. This involves both being able to control the AI, e.g.\,guiding it toward the correct decision, and making sense of the AI output  \cite{abdul2018trends}, particularly in instances when the AI's decision does not match that of the user.  Experimental evidence shows that people trust the AI more and adjust their own predictions to match the AI's if the model's accuracy is observed to be high, but are generally unlikely to do so merely on the basis of high stated accuracy \cite{yin2019understanding}.

Many statements on the importance of trust in AI are instead couched in \textit{instrumental} terms. It is commonly asserted that trust is critical for realizing the full economic and societal benefits afforded by AI, insofar as lack of trust will lead to lack of adoption across the economy  \cite{dietvorst2015algorithm,dutton2006trust,hengstler2016applied,lee2018understanding,sabater2005review,siau2018building}. While this argument may hold strongly for purchasers (a type of \textit{controller}), \textit{user} and \textit{subject} distrust may not be determinative of non-use \cite{Knowles:2018:OAD:3266364.3196490} due to various social and/or organizational entanglements with the technology that prohibit disuse. Other instrumental TAI arguments include usage continuance  \cite{dietvorst2015algorithm,dietvorst2016overcoming,kizilcec2016much,Lankton:2011:MTF:1989098.1989101,Lim:2009:WWE:1518701.1519023,hancock2011meta,Pu:2006:TBE:1111449.1111475,parasuraman1997humans}, permitting access to data for use within the AI \cite{woolley2019trust}, and promoting productivity and/or efficiency within organizations using AI  \cite{Passi:2018:TDS:3290265.3274405,vanneste2014trust} (see also \cite{Knowles:2014:TD:2531602.2531699,mcknight2000trust,riegelsberger2005mechanics}).

Less commonly, trust in AI is seen to be \textit{intrinsically} valuable. In sociological literature, trust is considered ``a functional prerequisite for the possibility of society'' \cite{lewis1985trust}, the foundation of all social relationships (see \cite{ullman2018does}); 
and it enables humans to take action on the basis of expectation despite the inherent unknowability of the future \cite{mollering2001nature} (see also \cite{Knowles:2015:MPT:2675133.2675154,Passi:2018:TDS:3290265.3274405}). Trust is not only an enabler of a functional society, it affords psychological benefits to its members, e.g.\,mitigating fear when facing vulnerability \cite{kramer1999trust}, feeling confident about one's actions despite uncertainty \cite{doshi2017towards,lipton2017doctor,siau2018building}, and experiencing an emotional bond \cite{toure2015or}. 
The \textit{intrinsic} view makes room for healthy distrust as a key to a harmonious relationship with AI: People may be willing to rely on AI while being aware of issues with the AI that pose a threat to this relationship (see \cite{mcknight2001trust}). Regardless of ultimate adoption, distrust tends to signal that ``important matters are at stake'' \cite{Knowles:2018:OAD:3266364.3196490} and, if taken seriously, can exert a corrective force on the development of AI which brings it into closer alignment with human values  (e.g.\,\cite{balaram2018artificial,Binns:2018:RHP:3173574.3173951,lepri2018fair,Veale:2018:FAD:3173574.3174014,Woodruff:2018:QEP:3173574.3174230}). It may be that, given limited agency in disusing AI, not to mention the challenges in developing ethical AI \cite{cave2018motivations}, distrust may in fact be the most appropriate ``mechanism for dealing with risk'' \cite{mcknight2001trust} entailed by the technology.

\subsection{Core pillars}
Figure 1 illustrates the complexity of TAI and the many different ways it might be approached. Assuming any given orientation to TAI is comprised of one category of agent, object, basis, objective, and impact,\footnote{In reality, it is common for papers to speak to multiple categories at once when exploring TAI.} and that none of these are inherently incompatible, there are potentially as many as 243 combinations, each representing a different facet of TAI. That is not even accounting for the many subcategories that have been abstracted away---notably, different types of \textit{user} (non-expert versus expert, in various contexts of use) or \textit{controller} (developer, purchaser, monitor, regulator), and of course, a huge variety of types of \textit{technology}. And yet, emerging from this ontology is a set of three fairly coherent core pillars within the TAI literature (see columns, bold vertical pathways). The first pillar, left on the diagram, is concerned with \textit{optimizing decision making}: One must understand enough of what the AI is doing and its particular limitations to be able to determine what action to take given its output. Transparency is the intuitive means of promoting appropriate reliance so that the agent maintains trust in the AI as a tool to inform their decision making. The second pillar, middle, is concerned with \textit{maximizing uptake}: Organizations embedding AI into their products and processes must gain the confidence of those affected by the AI so that they embrace it/them. Commonly, this is done through efforts to visibly attend to the matter of fairness. And the third pillar, right, is concerned with \textit{minimizing harm}: People must feel safe and comfortable in an increasingly AI-driven society. Widespread distrust of AI requires developing a system of structures, e.g.\,accountability mechanisms, that ensure that AI is trustworthy.

\section{Treatments of trust in FAT literature}
In the 2+ years since the curation of the above corpus, there has emerged a sufficiently large, focused collection of works on fairness, accountability, and transparency to analyze for treatments of trust. To create our second corpus, we searched all papers to date appearing in the ACM Conference  on Fairness, Accountability, and Transparency (formerly FAT*, now FAccT) proceedings for instances of ``trust''  within the abstract, keywords or body of the paper.\footnote{We recognize that other venues might have been included which have relevant contributions to FAT literature. Our analysis to follow is limited to the ACM FAccT community during the years 2018--2021. Future work could apply this ontology to compare FAT venues and/or look for changes to FAccT over time.} Papers were included if they used at least once the term ``trust'', ``distrust'', ``trustworthiness'', and/or ``trustability'' (excluding uses of these terms only in list of references, and excluding the legal term ``anti-trust'' and the phrase ``absolute trust'' in statistical measures \cite{taskesen2021statistical}); we also excluded papers that did not provide enough surrounding context when mentioning trust to derive any codes (only one paper: \cite{kasirzadeh2021use}), papers using trust only when describing study limitations \cite{celis2021effect}, and papers where the only mentions of trust were in reference to something other than AI (e.g.\,a mass spectrometer \cite{geiger2020garbage}). This resulted a total of 57 papers,\footnote{Two of these papers overlap between corpus 1 and corpus 2: \cite{Leong:2019:REW:3287560.3287591,mittelstadt2019explaining}.} or 24.3\% of the total FAT*/FAccT proceedings---a very significant proportion, indicating the centrality of trust and justifying its thorough treatment here.  Emulating the directed content analysis performed on the first corpus for this second corpus, we coded all papers according to  \textit{agent}, \textit{object}, \textit{basis}, \textit{objective}, and \textit{impact}. It is important to note that this was a non-trivial exercise. Codes needed to be inferred, as authors do not have a common language for talking about their particular orientation to trust, and indeed may not realize that their orientation may differ from others and thus need spelling out. This makes the ontology highly valuable, as it will enable authors to establish their position relative to a common frame.

Our analysis did not surface any additional categories unaccounted for in the ontology---itself an interesting finding given that this need not necessarily have been the case. There are in fact many similarities between the two corpora; in our summary of the FAT corpus below we focus on differences, highlighting new contributions from and particular emphases by the community. But beyond this, we hope to show that the ontology can be usefully applied to better understand individual papers and bodies of works. In section 3.6 we explore how the technical levers for promoting trust (i.e.\,fairness, accountability, and transparency) refract through this lens, and raise questions about potential anomalies.

\subsection{Trusting agents in FAT literature}

The FAT corpus contributes new (if limited and conflicting) data regarding potential human biases regarding AI. While one paper finds evidence of automation bias in high-impact decision making contexts, with students perceiving automated decision making (ADM) as fairer than human decision making (HDM) for university admissions \cite{marcinkowski2020implications}, another finds a lack of consensus whether ADM was fairer than a human judge for decisions relating to granting bail \cite{harrison2020empirical}. The corpus also shows continued interest in  (in)appropriate reliance \cite{zhang2020effect}, with concern regarding \textit{users'} susceptibility to disinformation \cite{mustafaraj2020case}, as well as new findings that although \textit{users} can ``somewhat differentiate correct machine predictions from incorrect ones,'' these instincts can be deceived by plausible yet spurious explanations \cite{lai2019human}. While some work focuses on trust by \textit{users} who are highly motivated to inspect the AI's trustworthiness \cite{card2019deep,panigutti2020doctor}, the dominant concern of corpus 1, other work questions the interest \textit{users} have to engage with the AI at all (e.g.\,doctors preferring to be patient-oriented~\cite{sendak2020human}) and the capacity of \textit{subjects} to engage in meaningful deliberations on AI's trustworthiness~\cite{knowles2021sanction}.

There is a particularly strong focus in this corpus on the trust of the \textit{subject}, including a detailed examination of public TAI as qualitatively different to TAI of \textit{users} \cite{knowles2021sanction}. Some works allude to \textit{users'} dual role as \textit{subject} of the effects of a data system  \cite{andrus2021we,stark2021ethics}. 
The focus on \textit{subjects} is further evident in works exploring the challenges of crafting  \textit{subject}-actionable explanations, contributions deepening understanding of when to use as well as how to create counterfactual explanations
\cite{barocas2020hidden,mittelstadt2019explaining,mothilal2020explaining} and offering other methods to facilitate algorithmic recourse \cite{karimi2021algorithmic,venkatasubramanian2020philosophical}. 
Compared to corpus 1, this corpus also shows a growing interest in the trust needs of \textit{controllers}, with new methods to help guide developers in developing and evaluating trustworthy AI  \cite{sokol2020explainability} and communicating trustworthiness to regulators \cite{slack2020fairness}.

A major contribution of FAT publications is an emphasis on trust being particular to a given agent and the challenges in both balancing \cite{malgieri2020concept} and tailoring trust-promoting efforts for multi-user \cite{hamon2021impossible,thornton2021fifty} and multi-stakeholder contexts \cite{sendak2020human,suriyakumar2021chasing,washington2020whose}. Works illustrate tensions between the \textit{user's} needs (e.g.\,to understand model performance) or \textit{controllers'} needs (e.g.\,to evaluate model fairness) and the needs of the \textit{subject}  (e.g.\,right to privacy) \cite{andrus2021we,bogen2020awareness,park2021designing,suriyakumar2021chasing}; the importance of attending to \textit{subjects'} diverse needs, with evidence of clear hierarchies of beneficiaries \cite{washington2020whose}; and the limitations in usefulness of explanations to \textit{users} compared to \textit{controllers} \cite{bhatt2020explainable}. This corpus clearly responds to the  call of corpus 1 to reconcile developers' and regulators' assessments of trustworthiness \cite{knowles2021sanction,kroll2021outlining,malgieri2020concept,slack2020fairness,sokol2020explainability}. While there is a growing interest in audits as a means of promoting trust~\cite{knowles2021sanction,roldan2020dirichlet,wilson2021building}, this corpus reiterates the concern that fairness constructs differ between agents~\cite{kasinidou2021agree,malgieri2020concept,slack2020fairness,wilson2021building}---e.g.\,what a \textit{controller} would consider `unbiased' is not the same as what a \textit{subject} would consider `fair' \cite{harrison2020empirical}. There is, furthermore, a concern that AI audits may promote unwarranted trust, with reliance on statistical fairness metrics flattening the discourse and discouraging deeper analysis of structural inequality~\cite{chasalow2021representativeness,green2020false,kulynych2020pots}.

\subsection{Objects of trust in FAT literature}

The majority of transparency and fairness contributions within the FAT literature are concerned with trust in the \textit{technology} itself (see Figure 2). This is accompanied, however, by critique that a focus on trust in the \textit{technology} perpetuates the problematic notion of model objectivity \cite{green2020false}; that, indeed, ``trust in a technology is rooted in relationships---not in a technical specification or feature'' \cite{sendak2020human}. We also see a further blending of trust in \textit{technology} and trust in the \textit{organization}, as seen with concern regarding management of untrustworthy data sources \cite{mustafaraj2020case}; discussion of how AI narrows socially constructed categories~\cite{hancox2021epistemic,miceli2021documenting}; and explorations of the trustworthiness of the human processes involved throughout the AI development pipeline \cite{raji2020closing}, including in problem formulation \cite{passi2019problem}, model development \cite{hutchinson2021towards}, data labeling \cite{geiger2020garbage}, and creation of explanations \cite{hancox2020robustness}.

A resounding call from the FAT corpus is for more situated understandings of trust within deployment contexts \cite{abebe2021narratives,barabas2020studying,sendak2020human} and even ``re-contextualising data and models'' \cite{sambasivan2021re}. \textit{Technology} as not a thing one can trust absent context, as it may be trustworthy (e.g.\,`fair') in one context but untrustworthy in others \cite{roldan2020dirichlet,slack2020fairness}; thus, new methods are needed for making boundary conditions interpretable \cite{slack2020fairness}. There is, further, an emphasis on the role of ``human infrastructures'' \cite{sambasivan2021re} in facilitating dialogue that  enriches understanding of context in ways that both improves trustworthiness and promotes interpersonal trust in \textit{organizations} (see also \cite{katell2020toward}). Perhaps most notable of this corpus is the discussion of the effect of power dynamics on (dis)trust in a given a context \cite{abebe2021narratives,barabas2020studying,chasalow2021representativeness,kulynych2020pots,sambasivan2021re}. Works emphasize the importance of empowering \textit{subjects} as beneficiaries of these data systems to mitigate distrust of \textit{organizations} \cite{abebe2021narratives,sambasivan2021re}, and fundamentally challenging the \textit{structural} contributors to decisional power asymmetries~\cite{barabas2020studying,katell2020toward,kulynych2020pots}.

This corpus demonstrates rising interest in promoting trust in \textit{structures} to compensate for low levels of trust in \textit{organizations'} commitment to mitigating AI harms \cite{kulynych2020pots}. Efforts include promoting accountability for AI systems \cite{kacianka2021designing,kroll2021outlining,raji2020closing,roldan2020dirichlet}; establishing clear processes for \textit{subjects} to challenge unfair decisions \cite{karimi2021algorithmic}; establishing robust intra-\textit{organizational} governance practices \cite{hutchinson2021towards,knowles2021sanction,raji2020closing,wilson2021building}; and empowering of a network of  \textit{controllers} (e.g.\,regulators, auditors, journalists, advocates) \cite{knowles2021sanction,metcalf2021algorithmic,ribeiro2020auditing,roldan2020dirichlet,sambasivan2021re,wilson2021building}. Interestingly, a number of publications characterize the \textit{technology} itself as \textit{structure}, insofar as the AI can contribute to systemic bias \cite{green2020false,suriyakumar2021chasing} and growing power imbalances \cite{abebe2021narratives} which contribute to distrust. The intermingling of these different objects of trust is evident in the effort made within 
the FAT corpus to attend in a more rounded way to trust in \textit{technology}, \textit{organizations}, and \textit{structures} \cite{abebe2021narratives,sambasivan2021re,sendak2020human}---in short, cultivating trust ``ecosystems'' \cite{knowles2021sanction,sambasivan2021re} that promote trustworthy \textit{technology} and impel trustworthy \textit{organizations}.

\subsection{Bases of trust in FAT literature}
As with the first corpus, a \textit{cognitive} view of trust dominates the transparency, interpretability, and explainability contributions, at least within the FAccT conference archives. And yet, these works reflect waning confidence in the usefulness of explanations to end-users---non-expert end-users rarely trust the explanations \cite{hancox2020robustness}, decision-making by expert end-users is not improved greatly by explanations \cite{jesus2021can}, and contrastive explanations that succeed in helping users understand prediction errors nonetheless do not improve trust in the model \cite{lucic2020does}. The community calls for the development of 
consistent criteria for evaluating explanations \cite{sokol2020explainability}, with resultant trust being a key consideration. The corpus clearly acknowledges a \textit{social} (i.e.\,interactional) component to successful explanations, with works exploring how to make explanations more intuitive so that they ring true \cite{patel2021high}, tailoring explanations to the informational needs of recipients \cite{sokol2020explainability} (including preserving subject autonomy \cite{barocas2020hidden}), and even asking whether explanation recipients can contribute to constructing  explanations that work for them~\cite{hancox2021epistemic}.

Notably, the FAT corpus places slightly more emphasis on \textit{social} trust than \textit{cognitive} trust, and proportionally much greater overall emphasis on \textit{social} aspects, perhaps reflecting the social science leanings within the FAT community. Two in-depth explorations of TAI within the corpus \cite{jacovi2021formalizing,toreini2020relationship} draw heavily from \textit{social} trust models---both seeing TAI as ``rooted in, but nevertheless not the same as, interpersonal trust as defined by sociologists'' \cite{jacovi2021formalizing}. The corpus raises a number of new \textit{social} considerations. One is that \textit{subjects'} belief in an \textit{organization's} cultural dissimilarity can reduce trust \cite{abebe2021narratives}. Not only does dissimilarity breed suspicion regarding the \textit{organization's} Benevolence \cite{abebe2021narratives} (as in the Ability-Benevolence-Integrity model), it may affect their Ability to deliver TAI. 
The work highlights the importance of embracing cultural differences in notions of fairness \cite{abebe2021narratives,harrison2020empirical,sambasivan2021re} (see also \cite{marcinkowski2020implications}); taking into account the on-the-ground social realities of \textit{users} and \textit{subjects} \cite{abebe2021narratives,green2020algorithmic,katell2020toward}; co-developing measures of risk with affected communities so that they reflect real harms people experience \cite{katell2020toward,kulynych2020pots,metcalf2021algorithmic}; and being seen to make good faith efforts to balance different stakeholder interests \cite{malgieri2020concept,marcinkowski2020implications}. 

The \textit{affective} dimension remains comparatively under-developed, despite allusions to unaccounted for dynamics limiting trust \textit{users} and \textit{subjects} have in both the \textit{technology} and the \textit{organization}. The lack of \textit{structural} guarantees of trustworthiness has been identified as creating conditions for \textit{affective} distrust to thrive \cite{knowles2021sanction}, as is the sense that one has no recourse to algorithmic harm \cite{venkatasubramanian2020philosophical} and that \textit{organizations} are all-powerful \cite{kacianka2021designing,washington2020whose}. A proposed counterweight is obligating  \textit{organizations} to act in the best interest of those affected by their AI, as in a fiduciary model \cite{barocas2020hidden}. The corpus is also greatly concerned with \textit{subjects} experiencing harm (more on this in section 3.5), demonstrating an understanding that feeling betrayed or insufficiently cared for diminishes trust in ways not easily recovered from \cite{abebe2021narratives}.

\subsection{Trust objective in FAT literature}
\textit{Maintaining} trust is less robustly linked with the transparency, interpretability, and explainability contributions in the FAT corpus compared to the first corpus. This may be a function of the particular kinds of agents these corpora focus on. \textit{Maintaining} trust is a key objective in relation to agents whose use is required and/or whose trust facilitates improved decision making outcomes \cite{jesus2021can,lai2019human,zhang2020effect}; whereas \textit{gaining} trust is more salient to those whose use is in doubt and/or whose trust is important for intrinsic reasons~ \cite{bhatt2020explainable,Leong:2019:REW:3287560.3287591,lucic2020does,panigutti2020doctor}.

By far the most common objective within the FAT corpus is \textit{gaining} trust; however, the literature reiterates the importance of cultivating understanding of when not to trust the \textit{technology}, 
proposing the terminology of ``calibrating'' trust as more apt \cite{zhang2020effect}. Indeed, a particularly noticeable feature of the FAT corpus is its exploration not of how fairness, accountability, and transparency promote trust, but how these either can or should promote \textit{distrust}. It is noted that highly usable visualization tools promote trust in the \textit{technology} by data scientists  without requiring them to understand how it works \cite{hancox2021epistemic}. There is a concern that efforts to improve representativeness within datasets can inspire trust, lending the AI ``rhetorical authority, merited or not'' \cite{chasalow2021representativeness}. It is also argued that AI may promote unwarranted trust in both \textit{organizations} and  \textit{structures}, as it  ``plays a particularly important role in political contexts rife with distrust, in which officials facing external scrutiny need to depoliticize their actions by `making decisions without seeming to decide''' \cite{green2020false}. Similarly, China's Social Credit System attempts to compensate for high levels of distrust in society through dubious attempts to quantify citizens' trustworthiness \cite{engelmann2019clear}.

\textit{Restoring} trust is a more common objective in this corpus than the first, perhaps resulting from greater attention to the matter of distrust. Two important contributions emerge. The first is that public distrust in \textit{organizations} is seen as contributing to tighter controls over access to demographic data, which limit \textit{controllers'} ability to evaluate the fairness of algorithms \cite{andrus2021we}. The second is to shine a spotlight on the need for particular care in \textit{restoring} trust in contexts where distrust results from legacies of colonialism, disempowerment, and/or discrimination \cite{abebe2021narratives,barabas2020studying,bogen2020awareness,green2020false,park2021designing,sambasivan2021re,suriyakumar2021chasing,washington2020whose}. 
\subsection{Impact of having trust in FAT literature}
The various FAT publications concerned with enabling agents to identify when the \textit{technology} is untrustworthy \cite{card2019deep,lai2019human,slack2020fairness,zhang2020effect} are clearly justified by the \textit{epistemic} rationale for TAI: Knowing when to trust AI is key to using it in ways that lead to improved decision making (see also \cite{card2019deep,karimi2021algorithmic,mittelstadt2019explaining}). (Interestingly, this argument is extended to collaborating data scientists who need to know which datasets and algorithms produced by which collaborator are trustworthy in order to better inform policymaking \cite{thornton2021fifty}.) A new, and particularly striking contribution of the corpus, however, is its problematization of trustworthiness. Works call for deeper examination of the ``\textit{epistemic} and methodological underpinnings of algorithms'' \cite{green2020false} (italics added;\footnote{We recognize that the term `epistemic' is used somewhat differently here to our meaning; however, the questions posed by this author nonetheless relate to our \textit{epistemic} category.} see also \cite{chasalow2021representativeness}), questioning whether algorithms are faithful enough representations of complex phenomena to derive trustworthy outputs \cite{hancox2021epistemic,passi2019problem} and whether explanations ``reflect real patterns in the data or the world'' \cite{hancox2020robustness}, how to determine when datasets are representative in a meaningful way \cite{chasalow2021representativeness}, and whether TAI principles can be implemented in a way that can be audited \cite{malgieri2020concept}.

As with the TAI corpus, we see the \textit{instrumental} rationale in (limited) works concerned with adoption \cite{lucic2020does}, as well as those noting that \textit{subjects'} distrust of \textit{organizations} presents a barrier to collecting data \cite{andrus2021we,sambasivan2021re}. As to the latter, attending to algorithmic fairness is thus seen as \textit{instrumentally} valuable in ``enabling data collection in low-trust environments'' \cite{andrus2021we}; whereas other works propose fairness as important for reputational preservation, including preserving trust in institutions (e.g.\,higher education) \cite{marcinkowski2020implications}. There remains an implicit concern that distrust will limit uptake of AI. To address this, AI audits have been proposed as a means of engendering trust, but now the FAT community is wrestling with how to instill trust in the audits themselves \cite{wilson2021building}.

Intrinsic rationales for TAI are much more prominent in this corpus than in the first corpus, perhaps because \textit{subject} trust, a strong focus in FAT literature, is not easily justified in \textit{instrumental} or \textit{epistemic} terms. It is clearly seen as ``fundamental to human flourishing'' that people have ``peace of mind'' \cite{venkatasubramanian2020philosophical} regarding the use of AI in society. FAT literature emphasizes that this trust will only be earned by effectively mitigating AI harms \cite{green2020algorithmic,katell2020toward,kulynych2020pots,malgieri2020concept} (i.e.\,real harms to real people~\cite{metcalf2021algorithmic}), while at the same time ensuring more equitable distribution of AI benefits \cite{washington2020whose}. While it is common for TAI to frame distrust as largely unwarranted (e.g.\,see \cite{cave2019scary} from the first corpus)---seeing this distrust as possibly arising from a failure to communicate the technology's trustworthiness---the FAT literature instead proposes distrust as the most appropriate stance towards AI at this time, given that potential harms and inequities have not been satisfactorily resolved.

\subsection{A bird's-eye view on FAT}

\begin{figure*}
  \includegraphics[width=1.0\textwidth]{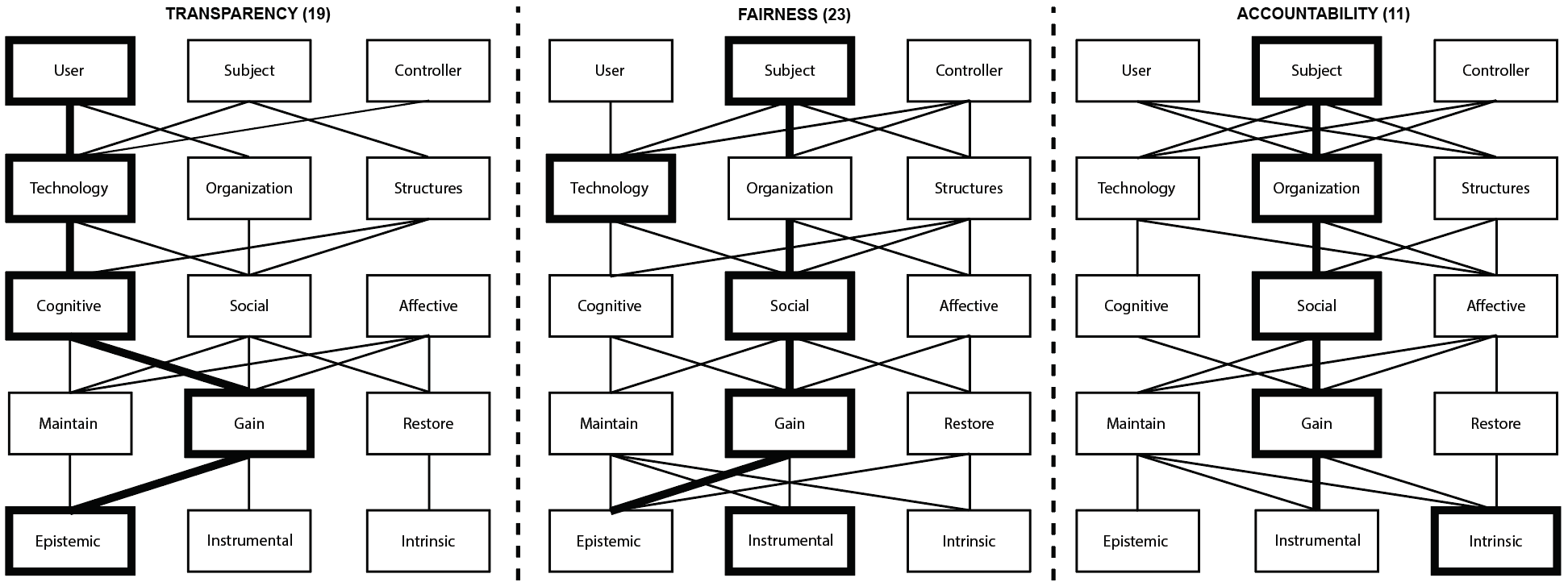}
  \caption{TAI ontology reflected through the FAT literature. Pathways indicate the presence of connections found in publications (i.e.\,at least one paper connecting the categories).   Bold pathways indicate   stronger (most common) connections between categories. Bold categories are the most commonly found. Number of publications per group is indicated by (X). NB: Four papers, which were in-depth treatments on trust, were not placed into groups above.}
\end{figure*}

Figure 2---breaking out the three promotion mechanisms individually---tells an interesting story about the FAT corpus. There are clearly very different orientations to trust between the transparency subset and fairness and accountability ones. And for the most part, these transparency publications conform to expectations, following down the left column the narrative of \emph{optimizing decision making}. It is surprising, however, that  here transparency is presented as a mechanism for \textit{gaining} trust more than for \textit{maintaining} it, particularly given that the TAI literature suggests that explanations are generally sought in moments when trustworthiness is suddenly thrown into question (i.e.\,when the AI does something unexpected).

The fairness literature also largely conforms to expectations, in line with the \textit{maximizing uptake} narrative (middle column), though it seems odd that a predominant object of trust in these papers is \textit{technology} 
(10 of 23 papers). Even if most often papers focus on \textit{subject} trust in an \textit{organization}, and some of the fairness papers which explore trust in \textit{technology} are concerned about \textit{user} or \textit{controller} trust, we have to question whether it makes sense to discuss \textit{subject} trust in \textit{technology}. Trustworthiness of the AI itself is largely beyond the capability of \textit{subjects} to gauge, and often they are not even in a position to question the trustworthiness of the AI because it acts upon them in ways unseen  \cite{knowles2021sanction}. Because of this, we might also have expected more of a recognition of the trust \textit{subjects} place in \textit{structures} as a means of ensuring fairness, i.e.\,freeing them from having to concern themselves with the AI's trustworthiness \cite{knowles2021sanction}. It also seems odd that while trust is seen as \textit{instrumentally} useful and the most common objective in the fairness works is to \textit{gain} trust (see bold boxes), these are rhetorically disconnected within individual publications, the most common connection being between \textit{gain} and \textit{epistemic} (see bold line).

The accountability subset diverges noticeably from the expected narrative (i.e.\,not proceeding down the right column), leaving us wondering if there is a coherent justification for accountability other than \textit{minimizing harm}. Granted, the accountability subset was the smallest of the three and perhaps more data are needed to see meaningful patterns; but why isn't the \textit{controller} a more important agent in these publications? Even among the ones looking at \textit{controllers'} trust, there is surprisingly little discussion in the accountability papers (at least the ones that mention ``trust'' that were included in this corpus) of trust by regulators. This is out of step with calls within the TAI literature for the need not only to regulate AI but, in order to do this effectively, to focus on what trustworthiness criteria regulators would look to (see section 2.1). For that matter, where is the focus on \textit{structures}? It seems odd for \textit{controllers} (if talking about regulators, that is) to trust in \textit{organizations}, particularly on the basis of \textit{social} trust, as opposed to trusting in an organization's internal governance \textit{structures}, for example their practices in producing algorithmic documentation \cite{raji2020closing}. Likewise, surely accountability mechanisms provide assurances of trustworthiness, and these assurances, the fact that it is governed by some capable entity, is what subjects would trust \cite{knowles2021sanction}. While we cannot answer these questions based on the current analysis, they may perhaps point to some of the confusions that make meaningful interpretation of `trust' difficult, and hence, to areas where the ontology proposed in this paper could help unify the narrative going forward.

\section{Conclusion}
In developing an ontology of TAI, it is tempting to consider whether it 
is a \textit{true} characterization of the literature. This, of course, is hard to say with certainty. The more interesting question, though, is whether it is \textit{useful} characterization. So what does this model do for us? 

The first thing it may do is reveal important distinctions that generally go unstated and therefore unacknowledged. Trust may be different enough between the myriad pathways through the model that it makes little sense to compare findings between papers following very different paths. Going forward, having a set of commonly understood referents may allow authors to make their orientations clear, helping to avoid issues of papers talking past each other and potentially muddling insights into how best (or even whether it makes sense) to promote trust in AI for a particular set of agents in a particular context. 

At the same time, despite its complexity, the ontology reveals strong connections between certain entities, helping to simplify what otherwise may seem like disjointed discourse. The three pillars are recognizable, if necessarily simplified, ideal-typical categories within TAI. While clearly not all TAI contributions fit within these three pillars, they may be described in contradistinction to these ideal types as a way of progressing the field.

And finally, we have shown that the ontology can be usefully applied as a lens. We were able to more clearly see what aspects of TAI the FAT community has been most interested in and where their contributions lie; see how contributions on fairness, accountability, and transparency differ in terms of their conceptualizations of trust; and even challenge whether certain arguments are coherent and certain mechanisms for promoting trust are appropriately conceived.

\section{Acknowledgments}

This work is partially funded by the ESRC funded grant BIAS:
Responsible AI for Labour Market Equality (ES/T012382/1) and
by the Data Science Institute at Lancaster University. The authors would like to thank the Partnership on AI for convening its expert group on Trust in AI, from which this collaboration sprang.

\balance{}

\bibliographystyle{ACM-Reference-Format}
\bibliography{sample-base}

\end{document}